\documentclass[conference]{IEEEtran}
\IEEEoverridecommandlockouts
\usepackage{cite}
\usepackage{amsmath,amssymb,amsfonts}
\usepackage{algorithmic}
\usepackage{graphicx}
\usepackage{textcomp}
\usepackage{xcolor}
\def\BibTeX{{\rm B\kern-.05em{\sc i\kern-.025em b}\kern-.08em
    T\kern-.1667em\lower.7ex\hbox{E}\kern-.125emX}}
\begin{document}

\title{Logic Solver Guided Directed Fuzzing for Hardware Designs

% \thanks{Identify applicable funding agency here. If none, delete this.}
}
\author{
    \IEEEauthorblockN{Raghul Saravanan and Sai Manoj P D}
    \IEEEauthorblockA{Department of Electrical and Computer Engineering \\
    George Mason University, Fairfax, VA, USA \\
    Email: \{rsaravan, spudukot\}@gmu.edu}
}

\maketitle

\begin{abstract} 

The ever-increasing complexity of design specifications for processors and intellectual property (IP) presents a formidable challenge for early bug detection in the modern IC design cycle. The recent advancements in hardware fuzzing have proven effective in detecting bugs in RTL designs of cutting-edge processors. The modern IC design flow involves incremental updates and modifications to the hardware designs necessitating rigorous verification and extending the overall verification period. To accelerate this process, directed fuzzing has emerged focusing on generating targeted stimuli for specific regions of the design, avoiding the need for exhaustive, full-scale verification.

%To address these growing challenges, recent advancements in hardware fuzzing have underscores its efficacy in detecting bugs in cutting-edge processors.  To circumvent the challenges, fuzzing works has introduced directed fuzzing—a focused verification strategy that emphasizes generating targeted input stimuli for specific regions of the design, rather than conducting exhaustive verification across the entire hardware design space. 

However, a significant limitation of these hardware fuzzers lies in their reliance on an equivalent SW model of the hardware which fails to capture intrinsic hardware characteristics. To circumvent the aforementioned challenges, this work introduces TargetFuzz, an innovative and scalable targeted hardware fuzzing mechanism. It leverages SAT-based techniques to focus on specific regions of the hardware design while operating at its native hardware abstraction level, ensuring a more precise and comprehensive verification process. We evaluated this approach across a diverse range of RTL designs for various IP cores. Our experimental results demonstrate its capability to effectively target and fuzz a broad spectrum of sites within these designs, showcasing its extensive coverage and precision in addressing targeted regions. TargetFuzz demonstrates its capability to effectively scale 30x greater in terms of handling target sites, achieving 100\% state coverage and 1.5x faster in terms of site coverage, and shows 90x improvement in target state coverage compared to Coverage-Guided Fuzzing, demonstrating its potential to advance the state-of-the-art in directed hardware fuzzing.

\end{abstract}


\section{Introduction} \label{intro}

Verifying hardware designs is an essential and complex phase in modern hardware design, ensuring system integrity and compliance with functional specifications. Over the years, the hardware verification community has developed various dynamic and formal verification techniques  \cite{Chen'11, Mukherjee'15}  through Electronic Design Automation (EDA) tools to mitigate the effects of the bugs. However,  both dynamic and formal verification techniques have failed to match the pace of ever-increasingly complex IC and System-on-Chips (SoCs) and are less efficient in terms of scalability \cite{Dessoky'19, Sarangi'06, Wagner'07}. As designs become more intricate, achieving high coverage becomes increasingly difficult due to the ever-expanding state space. The existing dynamic and formal techniques lack scalability, require significant human effort, and lead to exponentially increasing verification time \cite{Wang'18,Tiwari'11}.

To address the above-mentioned challenges, recently hardware community researchers introduced hardware fuzzing inspired by software testing paradigm \cite{saravanan2024fuzz,saravanan2025synfuzz,saravanan2024exploring,saravanan2024emergence}.  Fuzzing is a widely used software testing technique for bug detection in software applications \cite{microsoftrisk}. The existing works on hardware fuzzing  \cite{Kande'22,hybrid, Hur'21, Canakci'23} has proven its efficacy in bug-detecting capabilities in state-of-the-art open-source CPU RTL designs \cite{RISC-V, Morlkx}. Most of the hardware fuzzing works employ Coverage Directed Greybox Fuzzing (CGF), wherein fuzzing is steered by feedback mechanisms that guide test generation based on coverage metrics to explore new states and maximize code coverage. In CGF, new input seeds are generated by mutating interesting seeds—those that maximize coverage—to explore the uncovered regions of the RTL design, thereby maximizing code coverage. RFuzz \cite{Laeufer'18}, is one of the first hardware verification proposed employing CGF, to maximize the code coverage of RTL designs.

Analogous to software development, modern hardware development often involves iterative design updates and modifications, where only specific regions of the design are changed \cite{direct}. Traditional techniques, such as RFUZZ \cite{Laeufer'18} spend unnecessary resources and time maximizing code coverage for the entire design, rather than specifically targeting the modified regions. To accelerate the verification of modified (i.e., specifically targeted) regions of the design, DirectFuzz \cite{direct,saravanan2025intelligentgrayboxfuzzingatpgguided} was proposed. This approach introduced Directed Greybox Fuzzing (DGF), which aims to generate input seeds that maximize code coverage specifically for the targeted regions of the design. Unfortunately, DirectFuzz adopts translating the RTL design of the hardware to an equivalent software level of abstraction using tools like Verilator \cite{Verilator} and then fuzzing the resultant software code (i.e. fuzzing hardware as software model). In this case,  treating hardware as software fails to account for intrinsic hardware characteristics, which are crucial for accurate bug detection \cite{Kande'22}. %. Secondly,  hardware and software operate at different levels of abstraction, and
%unlike software, hardware does not inherently crash. In addition, one of the main challenges with these approaches is extracting suitable coverage metrics from the translated hardware model are not naturally aligned with the hardware’s unique
%behaviors \cite{}.

%As mentioned earlier, modern hardware designs come with an ever-expanding state space, making it both time-consuming and dependent on carefully chosen input seeds to reach specific states. These designs often capture data based on previous states, resulting in dependencies across multiple clock cycles. This complexity requires a significant number of input seeds to achieve an intended target state. 
% \textcolor{red}{Text above is not clear. Removed the above text!}

Furthermore, reaching a targeted state in the target regions is crucial for several reasons. The targeted states may relate to known failure modes, edge cases, or critical functional requirements essential for the correct operation of the hardware. By directing the input seed generation towards achieving the targeted states or gates in the target regions of the design, verification efforts become more focused and efficient, enabling deeper exploration of potentially critical areas. However, the existing DGF works do not reach the intended target state. Therefore, there is a compelling need for accelerated hardware fuzzing of targeted regions in the hardware design compatible with conventional industry-standard IC design and verification flow enabling seamless verification of hardware designs.

\textbf{Proposed Work and Key Contributions:} In this work, we propose a novel SAT-based targeted fuzzing approach, TargetFuzz, aimed at accelerating the verification of targeted regions in its innate hardware level of abstraction. This method leverages SAT solvers to generate input seeds specifically for the targeted regions of the design to enhance verification efficiency. \textit{
To the best of our knowledge, this is the first work to apply SAT solvers within the context of Directed Greybox Fuzzing (DGF)}. In addition,  TargetFuzz can be leveraged to reach a targeted state in the specific region of the design as detailed in Section \ref{targetstate}.
Moving forward, we propose to extract hardware coverage metrics for the target sites and targeted state coverage through industry-standard EDA tools. On the whole, our TargetFuzz offers the following benefits compared to DirectFuzz: 1) supports conventional hardware
design and verification methods 2) captures intrinsic hardware
characteristics using hardware coverage metrics 3) reaches the target sites and targeted states efficiently reducing verification time 4) is scalable for large and complex designs. The cardinal contributions of the proposed TargetFuzz are:
% outlined below:

\begin{itemize}
    \item We propose a novel fuzzer, TargetFuzz, that leverages SAT solvers to generate targeted input seeds aimed at maximizing coverage for the target regions of the hardware design at its native abstraction.
    \item Our technique can generate input seeds to reach the intended targeted states in the target regions of the hardware efficiently and faster.
    \item We present \textit{target state coverage} and \textit{target site coverage} metrics for the targeted fuzzing at its innate hardware level of abstraction through industry-standard EDA tools. 
    % \textcolor{red}{Are these metrics? Yes!}
\end{itemize}
To demonstrate the effectiveness of the TargetFuzz, we perform extensive evaluations on complex RTL designs. These include ISCAS benchmarks, PicoRV32 \cite{pico}, a RISC-V processor, as well as IP peripherals such as UART, AES, DSP, and sub-components of real-world open-source CPU designs (OpenRISC ISA)\cite{RISC} : 1) or1200 processor 2) mor1kx processor.

\section{Background}

In this section, we discuss the fundamentals of hardware fuzzing and the limitations of prior work.

\subsection{Background on HW Fuzzing} \label{background}

\begin{figure}[htb!]
% \vspace{-1em}
  \centering
  \includegraphics[width=0.9 \linewidth]{
  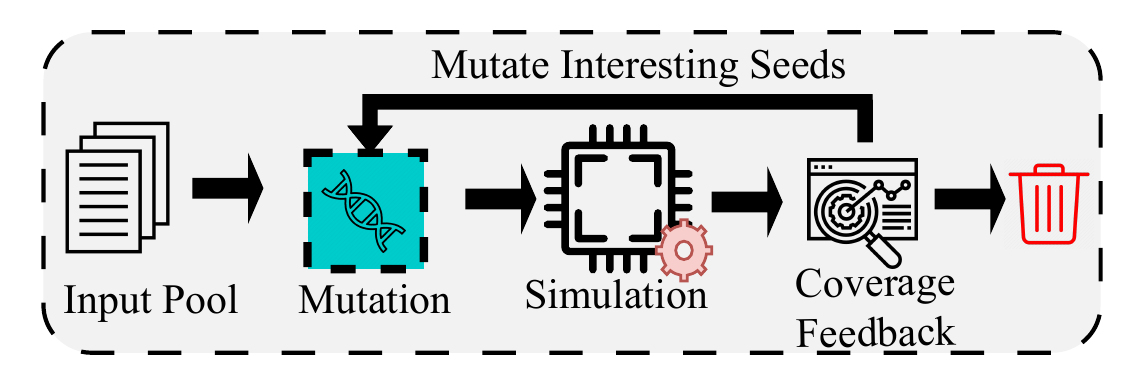} \vspace{-1em}
  \caption{Overview of Coverage Greybox Fuzzing (CGF)
  %\textcolor{red}{You have white space in fig, you need to remove it.}
  % \textcolor{blue}{acknowledged}
  % \cite{9} \textcolor{red}{Did you redraw this figure?\textcolor{violet}{Done}}
  } \label{fig:cgf}
 % \vspace{-0.5em}
\end{figure}

 To surpass the existing challenges in design verification, particularly in scalability for complex designs and automation, hardware fuzzing frameworks were proposed \cite{Kande'22,Trippel'22,hybrid}. The hardware concept of fuzzing primarily involves {\textit{1) random test case generation and mutations; 2) simulating the DUT (Design Under Test); and 3) analyzing for bugs or errors}} as illustrated in Figure \ref{fig:cgf}. The core concept behind traditional fuzzing begins with the generation of random acceptable test case generations (i.e., input stimuli). The DUT is simulated with these inputs to monitor its outcome, which is then verified against the expected output from the Golden Reference Model (GRM). However, when fuzzing large and complex designs, rather than relying solely on random inputs, a more efficient approach is to analyze the impact of these inputs on the DUT. Prior works have adopted code coverage as a feedback mechanism to the fuzzer (CGF), helping steer the fuzzer toward quickly exploring and covering the design space of the DUT as depicted in Figure \ref{fig:cgf}. 

On the contrary, Directed Greybox Fuzzing (DGF), is used for fuzzing at a particular region rather than the whole DUT, resulting in reduced verification time. Based on the chosen hardware level of abstraction and the fuzzing methodologies, the existing fuzzing frameworks can be classified into { 1) Direct Adoption of Software Fuzzer for Hardware \textcircled{A} 2) Fuzzing Hardware as Software \textcircled{B} 3) Fuzzing Hardware as Hardware \textcircled{C}.}  The existing works on DGF, such as DirectFuzz \cite{direct}, adopt \textcircled{B},  treating hardware as software fails to account for intrinsic hardware characteristics. Secondly,  hardware and software operate at different levels of abstraction, and unlike software, hardware does not inherently crash. In addition, one of the main challenges with these approaches is extracting suitable coverage metrics from the translated hardware model is challenging as these metrics are not naturally aligned with the hardware’s unique behaviors. 

To circumvent these challenges, fuzzing hardware as hardware \textbf{\textcircled{C}} was proposed \cite{Kande'22,hybrid,borkar2024whisperfuzz}, which retains the inherent properties of the hardware during the fuzzing process. This approach has proven effective in detecting bugs and vulnerabilities while preserving the hardware's native form, making it more suitable for complex hardware designs. However, while these methods adopt the CGF mechanism to increase coverage, our proposed TargetFuzz leverages targeted fuzzing to specifically target and fuzz hardware at the hardware level. Our TargetFuzz 
generates input seeds on reaching target sites and simultaneously 
reaches the targeted states directly, with less verification time.
% \textcolor{red}{Should this section be named as Background on HW fuzzing? Because prior work and related work seem confusing. Renamed !}

\begin{figure*}[h]
% \vspace{-1em}
  \centering
  \includegraphics[width=0.9 \linewidth]{
  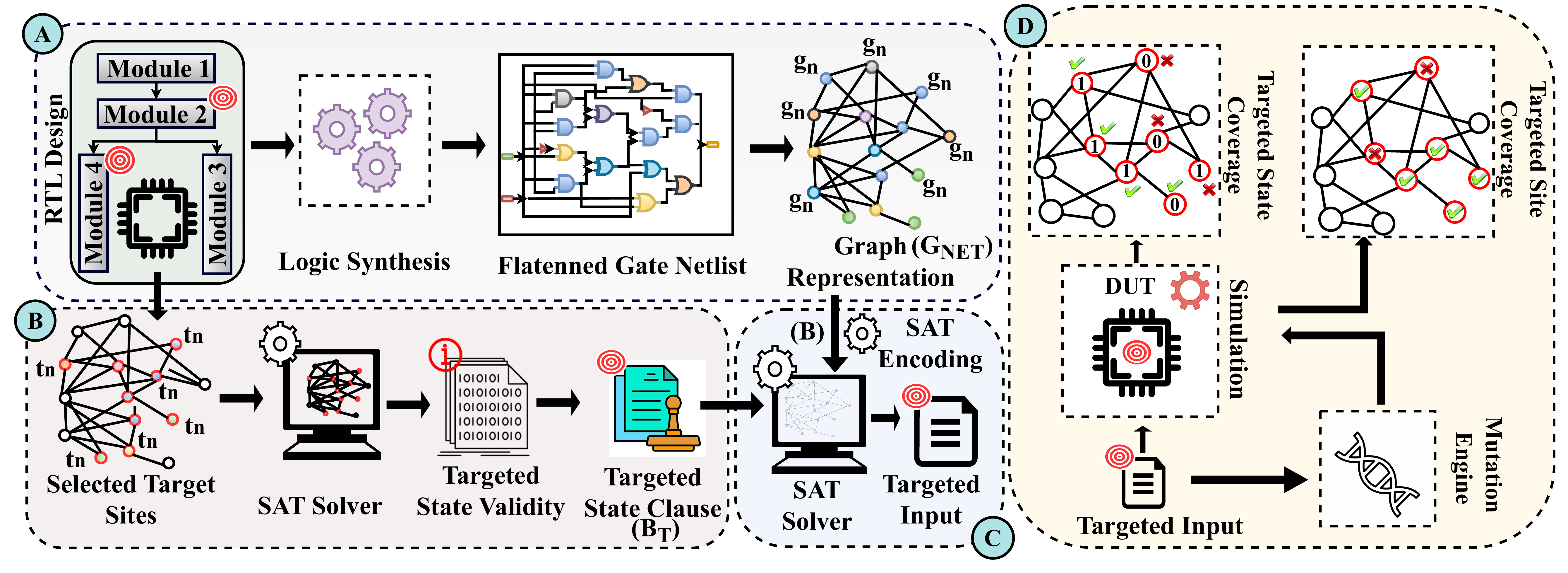} \vspace{-1em}
  \caption{Overview of TargetFuzz. \textcircled{A} Graph Generator and SAT Encoding, \textcircled{B} Target Site and State Selection, \textcircled{C} SAT-based Targeted Seed Generation, \textcircled{D} Targeted Fuzzing
  %\textcolor{red}{You have white space in fig, you need to remove it.}
  % \textcolor{blue}{acknowledged}
  % \cite{9} \textcolor{red}{Did you redraw this figure?\textcolor{violet}{Done}}
  } \label{fig:TargetFuzz}
 % \vspace{-0.5em}
\end{figure*}

%\subsection{Motivation} \label{motivation}

%Modern hardware designs are becoming increasingly complex, with an ever-expanding state space that makes reaching a specific state both time-consuming and reliant on carefully chosen input seeds. Reaching a targeted state in hardware verification is crucial for several reasons. Targeted states often represent specific points in the design where vulnerabilities or bugs are most likely to occur. These states can be associated with known failure modes, edge cases, or specific functional requirements that are essential for the intended operation of the hardware. By directing the system toward these states, verification efforts become more focused and efficient, allowing for deeper exploration of potentially problematic areas. In a typical sequential circuit, flip-flops capture data based on previous states, creating dependencies across multiple clock cycles. This requires complex input sequences to initialize the circuit into specific states. Our TargetFuzz can go and reach the targeted states directly, with fewer iterations from which the input can be mutated to explore further states. 

\subsection{Related Work}

To overcome the limitations in dynamic verification of RTL designs, where random inputs are applied to the DUT but often fail to detect deep-seated bugs, coverage-guided test generation was introduced \cite{Laeufer'18, Kande'22, hybrid, Trippel'22, Hur'21, direct, hyperprop}. Various coverage-guided test generation techniques for RTL designs have been proposed, inspired by fuzzing strategies from software testing. DirectFuzz \cite{direct} are notable examples that leverage directed fuzzing frameworks for more efficient RTL verification of specific sites. 

Furthermore, formal methods have been employed to generate test inputs for hardware designs, particularly to detect hardware Trojans in rarely exercised regions of the design \cite{det,Huang2016MERSST, scop}. Another interesting approach is concolic testing, also inspired by software testing \cite{Ahmed'18,Farahmandi, Meng'22}. Despite its potential, concolic testing in RTL designs faces scalability challenges due to the exponential growth of execution paths and complex state space in the hardware designs.

% \textcolor{red}{Stressing a bit more on DirectFuzz in the first paragraph will be helpful as our work can be compared with that.}

\section{Proposed TargetFuzz} \label{proposed}

To address the challenges associated with the existing DGF as outlined in Section \ref{background}, we present a novel SAT-based fuzzing framework, TargetFuzz for targeted fuzzing. Our TargetFuzz unfolds in four steps as shown in Figure \ref{fig:TargetFuzz}. Firstly, TargetFuzz generates the graph and SAT encoding i.e., conjunctive normal form (CNF) (Figure \ref{fig:TargetFuzz}\textcircled{A}) to perform SAT-based analysis on the RTL design (Section \ref{gsat}). Secondly, the target sites and their corresponding state selection (Figure \ref{fig:TargetFuzz}\textcircled{B}) are performed (Section \ref{sectargetsite}). Thirdly, TargetFuzz leverages SAT solvers for the targeted seed generation (Section \ref{seedgen}, Figure \ref{fig:TargetFuzz}\textcircled{C}) and finally the generated input patterns are used for Targeted Fuzzing (Section \ref{targeted fuzzing}, Figure \ref{fig:TargetFuzz}\textcircled{D}).  We detail the steps in the following subsections.

\subsection{Graph Generator and SAT Encoding} \label{gsat}

\textit{TargetFuzz is designed to generate input stimuli that target specific regions of an RTL design while maintaining operation at its native hardware abstraction level. }However, SAT solvers cannot directly process RTL code due to its high-level abstraction. The RTL designs of the DUT often consists of complex, interconnected modules and instances, which adds to the challenge. To bridge this gap, we first extract the flattened gate-level netlist via the EDA synthesis tool capturing the modules and instances from the original RTL design as shown in Figure \ref{fig:TargetFuzz}\textcircled{A}. This step ensures that the essential hardware characteristics are preserved, making the representation suitable for SAT-based analysis. 

By using the gate-level netlist, TargetFuzz can effectively leverage SAT solvers to perform targeted fuzzing while maintaining fidelity to the original hardware structure. The gate level netlist abstraction representing the RTL design subsequently is modeled as graph nodes where the propagation of signals from one gate to another is akin to passing information along the edges of the graph as shown in Figure \ref{fig:TargetFuzz}\textcircled{A}. The graph generator extracts the structural topologies of the netlist  and
represents the inputs, outputs, wires and gates of the netlist as graph nodes. From the graph nodes of the given netlist, the Boolean expression \textit{B} is generated by combining the logical equations of all individual gates and their interconnections. This can be represented as:

\begin{equation}
 B = \bigwedge_{i=1}^{N} G_i(I_i, O_i)
   \label{Bool} 
\end{equation}

where  \textit{N} is the total number of gates  in the netlist, $G_i$ represents the boolean equation of the \textit{i}th gate, \textit{$I_i$} and \textit{$O_i$} denote the inputs and the outputs of the \textit{$ith$} gate. The boolean equation \textit{B} is fed into the SAT solver as illustrated in Figure \ref{fig:TargetFuzz} during the targeted input stimuli generation outlined in Section \ref{seedgen}. The graph \textit{$G_{NET}$} with \textit{$g_n$} nodes 
representing the netlist, each graph node takes a value from \{0,1\}, with \textit{I} inputs. In the context of circuits, the values of these nodes represent the possible logical values of the netlist resulting from the input $I$. Each node $g_n$ in the graph \textit{$G_{NET}$} has an associated logical function $f(I,g_n)$, defined by the functionality of the RTL design. 
The inputs \textit{$I$} directly determine the possible state values of the graph nodes through the logic defined by the netlist, forming the 
universal valid graph state values \textit{U}. The universal valid graph state value \textit{U} encompasses all combinations of graph node values (\textit{v}) that result from the $2^{I}$ possible input states and the logic interactions defined by the RTL are formulated as:

\begin{equation}
    U=\{ (v_{g_1},v_{g_2},v_{g_3}.....,v_{g_n})  |   v_{g_j} \in  \text{\{0,1\}}\text{ } \forall g_j \in G_{NET}  \}
\end{equation}

where \textit{$G_{NET}$ =\{$g_1$,$g_2$, $\cdots$,$g_n$\}} is the set of all the nodes in the graph $G_{NET}$, \textit{$v_{g_j}$} represents the value of graph node \textit{$g_j$}, n is the total number of nodes and \{0,1\} indicates that each node \textit{$g_j$} can take a binary value defining the universal graph state values \textit{U}. Given that the RTL design is now represented as Boolean equations and the universal graph state values have been defined, the next step involves identifying the target sites and determining the corresponding target states at these selected sites as detailed in Section \ref{sectargetsite}.

\subsection{Target Site and State Selection} \label{sectargetsite}
 As we detail in Section \ref{targeted fuzzing}, the coverage points for TargetFuzz are targeted states of the target sites in the RTL design. To effectively guide input seed generation toward specific targeted sites in the RTL and achieve the desired target states within those sites, the process of target site selection and state selection is crucial as outlined in this section.

\subsubsection{Target Site Selection}TargetFuzz accommodates a wide range of target sites across various modules and instances. These target sites can reside within a specific module or instance, or span between two blocks, depending on the area of interest selected for targeted fuzzing. For the selection of target nodes, the target labels are derived from the RTL design and are inherently represented in the gate-level netlist, as illustrated in Figure \ref{fig:TargetFuzz} \textcircled{B} (target nodes are highlighted in red). Additionally, internal wires introduced at the gate-level netlist can also be considered as potential target sites, providing flexibility in the selection process to focus on specific internal circuit points for targeted analysis. This process can be manually selected or through an automated process. In the case of manual selection, verification engineers may choose nodes based on specific verification goals, such as areas prone to bugs or performance-critical sections or SCOAP scores \cite{scop}. For the latter, one could identify modified nodes by comparing the original graph structure with the modified graph structure to extract the nodes that have changed. We formulate the target sites in the RTL design as a set of target nodes $T$, shown in Equation \eqref{targetsite}

\begin{equation}
    T=\{ t_1,t_2....,t_k \}, t_i \in \text{$G_{NET}$ }  \forall i=1,2,...k
    \label{targetsite}
\end{equation}

where $t_i$ represents the individual target node and k is the total number of target nodes.

\subsubsection{Target State Selection} 
The early phase of Target State (\textit{$T_s$}) selection begins with defining the intended targeted state values of the target regions 
for the RTL design. The set of target nodes and their intended target states be defined as in Equation \ref{targetstate}, where $t_i$ represents a 
target node and $v_{t_i}$ $\in$  \{0,1\}  is the targeted state for the node $t_i$.

\begin{equation}
    T_s= \{ (t_1,v_{t_1}),(t_2,v_{t_2}),\cdots,(t_i,v_{t_i}) \}
    \label{targetstate}
\end{equation}
\vspace{-1em}

\begin{equation}
    T_s =
\begin{cases} 
\text{valid}, & \text{if } (v_{t_1}, v_{t_2}, \ldots, v_{t_i}) \in U \\
\text{invalid}, & \text{if } (v_{t_1}, v_{t_2}, \ldots, v_{t_i}) \notin U
\end{cases}
\label{valid}
\end{equation}

The targeted state information $T_s$ of the RTL design serves as the basis for the SAT solver to generate targeted input seed $I_T$. To ensure that there exists a targeted input seed $I_T$ $\in$ $I$ satisfying the targeted states $T_s$ simultaneously, we perform the \textit{targeted state validity} of the design as outlined in 
Figure \ref{fig:TargetFuzz}. The given targeted state $T_s$ specified by the verification engineer is considered valid if the state values at the target nodes belong to the universal graph state values \textit{U} as outlined in Equation \ref{valid}. The targeted state validity steers the SAT-based input generation process toward achieving a valid targeted state. It also ensures that when the target input $I_T$ is applied to the RTL design, the target regions ($t_i$) reach the specified state ($v_{t_i}$). It serves as a feasibility check before the boolean expression (\textit{$B_T$}) for the targeted state is derived. The validity of the targeted state \textit{$T_s$} for the given set of target sites \textit{T}  can be validated through Boolean SAT solver as outlined in Figure \ref{fig:TargetFuzz} \textcircled{B}. Moving forward, the boolean expression \textit{$B_T$} for the target sites for the RTL design with their corresponding targeted states is defined as :

\begin{equation}
    B_T= \bigwedge_{i=1}^{k} (f(t_i) = v_{t_i})                    
\end{equation}

where \textit{$f(t_i)$} is the boolean function for each target node \textit{$t_i$} and \textit{$v_{t_i}$} $\in$  \{0,1\} is the desired state 
for \textit{$t_i$}. For example, if the targeted nodes are ($t_1$,$t_2$,$t_3$) are their targeted states \textit{($v_1,v_2,v_3$)=(1,0,1)}, the boolean expression of the targeted states are defined as 
$B_T= \{(t_1)\wedge(!t_2)\wedge (t_3)\}$. Now that the target sites and their corresponding 
targeted states are encoded as boolean expressions, these can be fed into the Boolean SAT solver for the targeted input seed generation as outlined in Section \ref{seedgen} 

\begin{table*}[h!]

    \centering
    \caption{ {Experimental Results of TargetFuzz} }
    \vspace{-1em}
    \label{tb2:coverage_metrics}
    \scalebox{0.7}{
    
    \begin{tabular}{|l|c|c|c|c|c|c|c|c|c|c|}
    \hline  
    \textbf{Designs} & \textbf{Total \# of Gates}  &  \textbf{Total \# of Nodes}   &   \textbf{Total \# of Input Nodes}  &  \textbf{\% of Target Nodes }   &  \textbf{\# of Patterns} & \textbf{Time} & \textbf{State Coverage} & \textbf{Site Coverage} & \textbf{$D_{max}$} \\
    & & ($g_n$) & ($I$) & ($T$)  & ($I_T$)& ($s$)& ($T_C$)& & \\
    \hline
    C7552 &  3988  & 3968  &  207 & 6.3\%  &  100 & 0.12 &  100\% & 100\% & 9\\
     \hline
    C6288 & 5541   &  690  &  32  &  36.2\%  & 10  & 0.012 & 100\%  & 100\% & 1\\
     \hline
    S5378  &  1062  &  2603  &  35  &  28.8\%  & 66  & 0.06 &  100\% & 100\%& 15\\
    \hline
    S13207 & 1044 & 927 & 62 &  68.7\% & 85  & 0.10 &  100\% & 100\%& 12\\
     \hline
    UART & 556 & 945 & 29 & 38.32\%  & 17 & 0.034 & 100\% & 100\%& 10\\
     \hline
    or1200 IC FSM & 204 & 381 & 40 & 45.89\%  & 73  & 0.087 &  100\% & 90.13\%& 23\\
     \hline
    or1200 CTRL & 3726 & 4432 & 109 & 83.48\%  & 100  & 0.17 &  100\% & 94.78\%& 18\\
     \hline
    
    or1200 ALU & 6585 & 8079 & 123  & 43.8\%  & 52 & 0.07 &  100\% & 88.78\%& 3 \\
     \hline
    IBEX ALU & 3378   &  4414  &  202  & 74.7\%    & 5 & 0.03 &  100\% & 93.54\%& 7 \\
     \hline
    DSP & 30897 & 32261 & 34 & 89.2\% & 87 &  1.42 &  100\% &76.87\% & 27\\
    
    \hline
   AES  &   159175  & 162583    &  256  & 28.2\%   & 100  & 2.73 &  100\% & 81.28\% & 128 \\
    \hline
    PicoRV32  &  30149  &  36578  &  102  & 67.16\%   & 43  & 1.68 &  100\% & 73.77\%& 58\\
    \hline
    \end{tabular}
    }
   
\end{table*}

\subsection{SAT-based Targeted Seed Generation} \label{seedgen} 
The SAT solver is fed with Boolean expressions $B$ and $B_T$ to generate the targeted input pattern $I_T$ required for the fuzzing illustrated in Figure \ref{fig:TargetFuzz}\textcircled{C}. To enable comprehensive fuzzing on the targeted sites, the SAT solver can be invoked R times, where R is the required number of test patterns to be generated. Furthermore, to promote a broader exploration of the targeted input space $I_T$ and prevent redundancy in the input patterns, TargetFuzz incorporates the calculation of minimum hamming distance $D_{min}$ between any two generated target input patterns. This is formally represented in the Equations \eqref{boo}, and \eqref{hamminng}. The ($D_{min}$) ensures that there is no redundancy between the generated patterns. %and the latter ($D_{max}$) ensures to maintain the practical variability within the input space.

\begin{equation}
    I_T=SAT((B \wedge B_T)=1)  \text{ with } D_{min} \leq d(I^{(i)},I^{(j)}) \leq D_{max}
    \label{boo}
\end{equation}

\begin{equation}
    D_{min} > 1 \text{ and } D_{max} \leq I 
    \label{hamminng}
\end{equation}

\subsection{Targeted Fuzzing} \label{targeted fuzzing}

The RTL design is fuzzed with the targeted input $I_T$ which is generated by the SAT solver to fuzz the specific regions of the RTL design and reach the desired targeted state as shown in Figure \ref{fig:TargetFuzz} \textcircled{D}. TargetFuzz aids in targeted state exploration in the targeted sites impacted by the incremental changes or the regions that are likely to exhibit vulnerabilities or unexpected behavior when driven to their specified states. For instance, the bug within a particular region of the RTL design is known, these states can be incorporated into the targeted states $T_s$ for the SAT problem formulation, to explore the potential input space  $I_T$ $\in$ $I$ that can trigger the bug. 

For our TargetFuzz, the coverage points are the target sites of the design as shown in Figure \ref{fig:TargetFuzz} \textcircled{D}. While applying the targeted input seed $I_T$ to simulate the RTL design, the coverage metrics for the targeted region are extracted through the \textit{target site coverage} and \textit{target state coverage}. The \textit{target site coverage} is extracted through industry-standard hardware EDA tools which offer innate hardware coverage metrics for the target sites.  The target state coverage is a metric that is proposed in this work, to evaluate whether the specified states for the target regions have been reached as illustrated in Figure \ref{fig:TargetFuzz} \textcircled{D}\footnote{Nodes that reach the targeted state logic value are highlighted in green tick, while those that do not are highlighted in red tick. The values inside the node represent the logic values of the node, not the Coverage ($t_i$)}.  The \textit{target state coverage } ($T_C$) is defined as the ratio of the number of target nodes $t_i$ that have reached the targeted state $v_i$ to the total number of target nodes $t_i$. Given a set of target nodes $T$ in equation \ref{targetsite} with the targeted states in equation \ref{targetstate}, the targeted state coverage is defined as:

%In addition, TargetFuzz also leverages an option to mutate these seeds furthermore to explore the states other than the targeted states in the target regions as shown in Figure \ref{fig:TargetFuzz} \textcircled{D}.
%For our TargetFuzz, the coverage points are the states of the target sites in the RTL design.

\begin{equation}
\text{$Coverage$}(t_i)=
    \begin{cases}
        1, & \text{if } t_i \text{ has reached its intended state } v_{t_i} \\
0, & \text{otherwise}
    \end{cases}
\end{equation}

The overall target state coverage ($T_C$) for the the regions in T can be expressed as:

\begin{equation}
    T_C = \frac{\sum_{i=1}^{k} \text{Coverage}(t_i)}{k} \times 100\%
    \label{overallcoverage}
\end{equation}

\section{Experimental Evaluation}

\subsection{Experimental Setup}The proposed TargetFuzz is compatible with the conventional industry
standard IC design and verification flow enabling seamless targeted fuzzing of RTL designs. 
We leverage the industry-standard EDA tool, Cadence Genus, to generate the gate-level netlist
from a given RTL design. To extract the structural information from netlist for the graph generation and for deriving the boolean equations we use custom Python scripts. We used the SAT solvers provided by the \textit{PySAT} library. For the simulations of the RTL design with the generated input patterns we use, Cadence Xcelium, a standard EDA tool widely used in the industry and academia. To extract the target state coverage and target site coverage metric we leverage Cadence Integrated Metrics Center (IMC) through custom TCL and Python scripts. We conducted all our experiments on an Intel Core i7 processor (2.4 GHz) running on a Linux operating system.

We evaluate our TargetFuzz on varied combinational and sequential RTL designs including ISCAS benchmarks, UART \cite{Shakya2017BenchmarkingOH}, DSP block \cite{opentitan}, AES \cite{Shakya2017BenchmarkingOH}, ALUs of IBEX \cite{ibex}, or1200 Instruction cache controller and Control unit of or1200 \cite{RISC}. To demonstrate the scalability we also experiment on PicoRV32, a RISC-V processor from the OpenCores \cite{open}. For all the sequential circuits we assume full scan access \cite{Huang2016MERSST, det }. We set \textit{R}, the number of test patterns to generate as 100. However, these parameters can be tweaked based on the verification and user requirements.

\begin{figure*}[h]
% \vspace{-1em}
  \centering
  \includegraphics[width=0.8\linewidth]{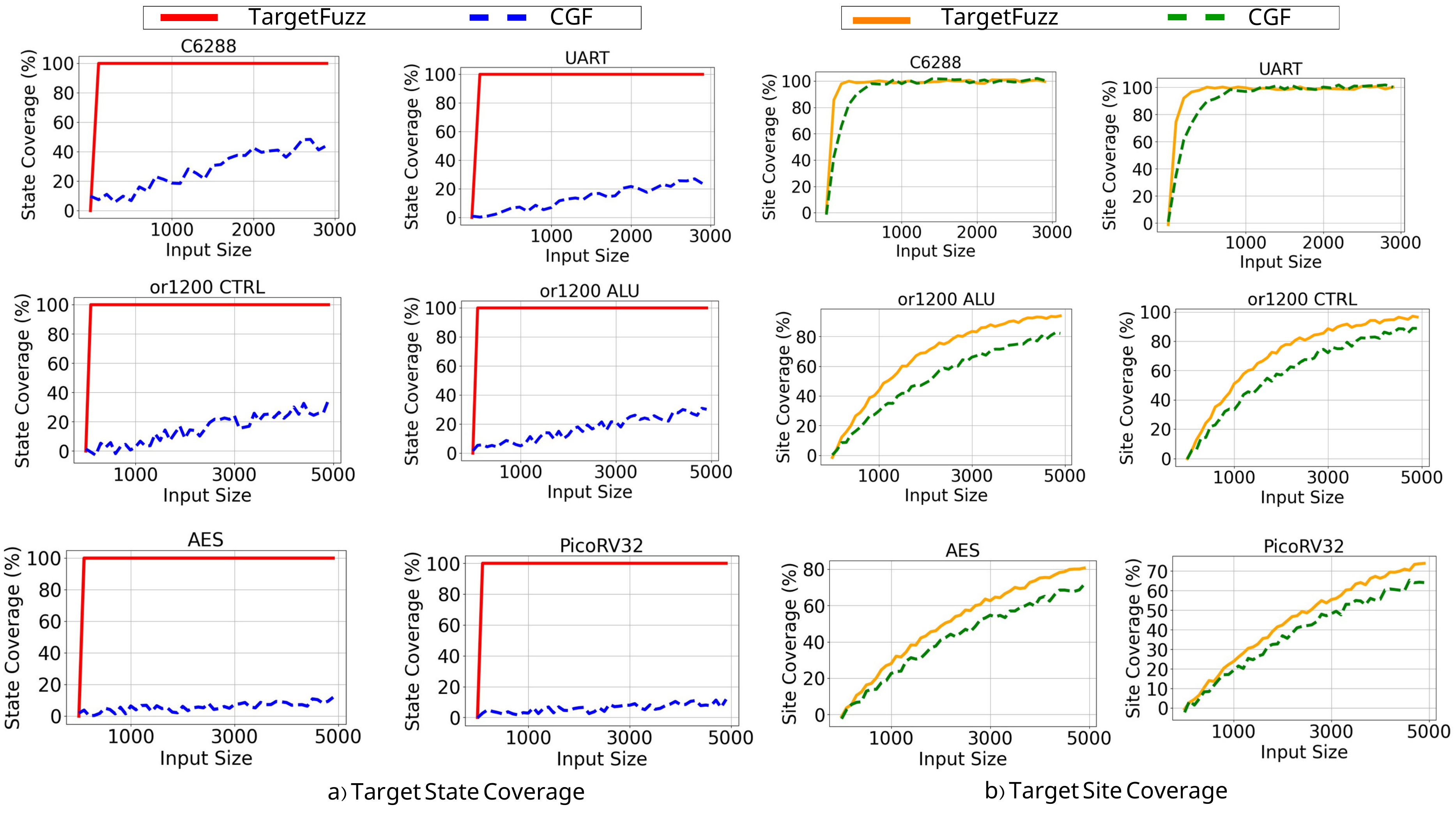}
%   images/satcov.jpg} \vspace{-1em}
  \caption{Coverage Results of TargetFuzz a) Target State Coverage b) Target Site Coverage
  %\textcolor{red}{You have white space in fig, you need to remove it.}
  % \textcolor{blue}{acknowledged}
  % \cite{9} \textcolor{red}{Did you redraw this figure?\textcolor{violet}{Done}}
  } \label{fig:satcov}
 % \vspace{-0.5em}
\end{figure*}

\subsection{Simulation Results}
In this subsection, we present the efficiency of our TargetFuzz where the goal is to reduce the testing time to reach targeted states in the target regions in the RTL designs. The experimental results of TargetFuzz are outlined in Table \ref{tb2:coverage_metrics}. The first column lists the hardware design names, followed by columns representing the total number of gates, the total number of nodes ($g_n$), and the total number of input nodes ($I$) for each design, respectively. The fifth column indicates the percentage of nodes selected from each design. The sixth and seventh columns represent the number of targeted input patterns ($I_T$) generated by the SAT solver and the time taken to generate those patterns, respectively. The eighth column shows the target state coverage ($T_C$) achieved, the ninth represents the coverage for the target sites and finally, the tenth column presents the highest Hamming distance found between the SAT-generated targeted patterns.

As outlined in Table \ref{tb2:coverage_metrics}, TargetFuzz achieves the lowest execution time for the C6288 design, while the highest execution time is observed in the AES design. The number of targeted input patterns $I_T$ found by TargetFuzz  depends on the design space and the proportion of the design selected for targeted fuzzing. Although PicoRV32 is smaller than AES and has a higher percentage of target nodes, TargetFuzz generates more test patterns for AES due to its extensive design space, which provides greater flexibility for the SAT solver. Conversely, while DSP is smaller than PicoRV32 but has more target nodes, TargetFuzz still generates more patterns for DSP highlighting that design complexity and target node distribution significantly impact pattern generation.

The number of patterns generated also depends on the proximity of the targeted nodes. Whether in smaller or larger designs, when the targeted nodes are closely located, the constraints are higher, resulting in fewer patterns. In contrast, when the nodes are more dispersed, TargetFuzz has higher explorability to explore the state space, leading to an increase in generated patterns. As the design space expands and the percentage of target nodes increases, the time taken to generate input patterns also rises. However, even a slight speedup in larger designs can save significant time in complex verification processes. In our benchmarks, for the largest designs such as AES and PicoRV32, with an average of 
47.68\% nodes selected, TargetFuzz was able to generate patterns within a few seconds achieving a remarkable targeted state coverage of 100\% as in Figure \ref{fig:satcov}.

The number of patterns generated and the time taken by TargetFuzz depend on several factors, including the design space, the number of target nodes selected, and the proximity of these nodes. On average, for each design, the time taken by TargetFuzz to generate a single pattern is approximately 0.008s. Overall, across all designs, TargetFuzz takes an average of 0.5 seconds to generate 61 targeted input patterns when the percentage of target regions is 50\% with a state coverage of 100\%. Furthermore, the coverage for the target sites is notable, with an average of 98.9\% acorss all the designs.

\subsection{Comparison with State-of-the-Art}
In this section, we present the coverage reports\footnote{
Due to space constraints, we have selected only a few designs for illustration purposes in the Figure \ref{fig:satcov}.} for the target state (Figure \ref{fig:satcov}a) and the sites (Figure \ref{fig:satcov}b) comparing TargetFuzz with CGF at its innate hardware abstraction, similar to \cite{Laeufer'18}. Due to the probabilistic nature of CGF, we ran our experiments fifteen times and reported the mean in Figure \ref{fig:satcov}. 

The results indicate that TargetFuzz consistently outperforms conventional coverage-based feedback fuzzing. TargetFuzz achieves 100\% targeted state coverage across nearly all tested designs, including complex architectures such as AES, DSP, and PicoRV32. Notably, TargetFuzz reaches high state coverage with significantly fewer input test patterns compared to the CGF approach. For complex designs such as AES, DSP, and PicoRV32, Coverage-guided fuzzing often achieves target state coverage of less than 10\%. Moreover, TargetFuzz achieves faster coverage in the target sites using its generated inputs, requiring fewer input patterns, as illustrated in Figure \ref{fig:satcov}(b) when compared to the CGF. On average, TargetFuzz is 1.5$\times$ faster than CGF in achieving target site coverage, with an average increase of 8\% in target site coverage. Additionally, TargetFuzz shows a 90$\times$ improvement in target state coverage compared to CGF.

\section{Conclusion}

This work introduces TargetFuzz, an innovative directed fuzzing mechanism designed to generate test input seeds for targeted sites in RTL designs, thereby accelerating hardware verification. Unlike prior approaches, TargetFuzz focuses on producing test patterns that achieve a specified targeted state at designated sites within the hardware design. Moreover, TargetFuzz operates at the native hardware abstraction level, rather than relying on software models of hardware as seen in other methodologies. 
The target state and site coverage metrics proposed in this work are specifically designed to align with the native hardware abstraction level, providing a comprehensive assessment of coverage that accurately represents hardware behavior. Experimental results on diverse hardware designs demonstrate  TargetFuzz is scalable and efficient for complex designs. TargetFuzz achieves 1.5x faster coverage than CGF in terms of target site coverage and achieving 100\%  target state coverage outperforming  conventional coverage-based fuzzing (CGF).

\bibliographystyle{IEEEtran}

\bibliography{ref}

% Generated by IEEEtran.bst, version: 1.14 (2015/08/26)
\begin{thebibliography}{10}
\providecommand{\url}[1]{#1}
\csname url@samestyle\endcsname
\providecommand{\newblock}{\relax}
\providecommand{\bibinfo}[2]{#2}
\providecommand{\BIBentrySTDinterwordspacing}{\spaceskip=0pt\relax}
\providecommand{\BIBentryALTinterwordstretchfactor}{4}
\providecommand{\BIBentryALTinterwordspacing}{\spaceskip=\fontdimen2\font plus
\BIBentryALTinterwordstretchfactor\fontdimen3\font minus \fontdimen4\font\relax}
\providecommand{\BIBforeignlanguage}[2]{{%
\expandafter\ifx\csname l@#1\endcsname\relax
\typeout{** WARNING: IEEEtran.bst: No hyphenation pattern has been}%
\typeout{** loaded for the language `#1'. Using the pattern for}%
\typeout{** the default language instead.}%
\else
\language=\csname l@#1\endcsname
\fi
#2}}
\providecommand{\BIBdecl}{\relax}
\BIBdecl

\bibitem{Chen'11}
M.~Chen and P.~Mishra, ``Property learning techniques for efficient generation of directed tests,'' \emph{IEEE Transactions on Computers}, vol.~60, no.~6, pp. 852--864, Feb 2011.

\bibitem{Mukherjee'15}
R.~Mukherjee, D.~Kroening, and T.~Melham, ``Hardware verification using software analyzers,'' in \emph{IEEE Computer Society Annual Symposium on VLSI}, 2015.

\bibitem{Dessoky'19}
G.~Dessouky, D.~Gens, P.~Haney, G.~Persyn, A.~Kanuparthi, H.~Khattri, J.~M. Fung, A.-R. Sadeghi, and J.~Rajendran, ``Hardfails: Insights into software-exploitable hardware bugs,'' in \emph{USENIX Conference on Security Symposium}, 2019.

\bibitem{Sarangi'06}
S.~R. Sarangi, A.~Tiwari, and J.~Torrellas, ``Phoenix: Detecting and recovering from permanent processor design bugs with programmable hardware,'' in \emph{IEEE/ACM International Symposium on Microarchitecture}, 2006.

\bibitem{Wagner'07}
I.~Wagner and V.~Bertacco, ``Engineering trust with semantic guardians,'' in \emph{Design, Automation \& Test in Europe Conference \& Exhibition}, 2007.

\bibitem{Wang'18}
F.~Wang, H.~Zhu, P.~Popli, Y.~Xiao, P.~Bodgan, and S.~Nazarian, ``Accelerating coverage directed test generation for functional verification: A neural network-based framework,'' in \emph{Great Lakes Symposium on VLSI}, 2018.

\bibitem{Tiwari'11}
M.~Tiwari, J.~K. Oberg, X.~Li, J.~Valamehr, T.~Levin, B.~Hardekopf, R.~Kastner, F.~T. Chong, and T.~Sherwood, ``Crafting a usable microkernel, processor, and {I/O} system with strict and provable information flow security,'' in \emph{International Symposium on Computer Architecture (ISCA)}, 2011.

\bibitem{saravanan2024fuzz}
R.~Saravanan and S.~M. Pudukotai~Dinakarrao, ``The fuzz odyssey: A survey on hardware fuzzing frameworks for hardware design verification,'' in \emph{Proceedings of the Great Lakes Symposium on VLSI 2024}, 2024, pp. 192--197.

\bibitem{saravanan2025synfuzz}
\BIBentryALTinterwordspacing
R.~Saravanan, S.~Paria, A.~Dasgupta, V.~N. Patnala, S.~Bhunia, and P.~D. Sai~Manoj, ``Synfuzz: Leveraging fuzzing of netlist to detect synthesis bugs,'' \emph{arXiv preprint arXiv:2504.18812}, 2025. [Online]. Available: \url{Paper=https://mason.gmu.edu/~rsaravan/papers/synfuzz.pdf URL=https://mason.gmu.edu/~rsaravan/projects/synfuzz.html}
\BIBentrySTDinterwordspacing

\bibitem{saravanan2024exploring}
R.~Saravanan and S.~M. Pudukotai~Dinakarrao, ``Exploring coverage metrics in hardware fuzzing: A comprehensive analysis,'' in \emph{Proceedings of the Great Lakes Symposium on VLSI 2024}, 2024, pp. 240--245.

\bibitem{saravanan2024emergence}
R.~Saravanan and S.~M.~P. Dinakarrao, ``The emergence of hardware fuzzing: A critical review of its significance,'' \emph{arXiv preprint arXiv:2403.12812}, 2024.

\bibitem{microsoftrisk}
\BIBentryALTinterwordspacing
Microsoft, ``Microsoft security risk detection,'' last Accessed : 6/1/2024. [Online]. Available: \url{https://www.microsoft.com/en-us/research/project/project-springfield/}
\BIBentrySTDinterwordspacing

\bibitem{Kande'22}
R.~Kande, A.~Crump, G.~Persyn, P.~Jauernig, A.-R. Sadeghi, A.~Tyagi, and J.~Rajendran, ``{TheHuzz}: Instruction fuzzing of processors using {Golden-Reference} models for finding {Software-Exploitable} vulnerabilities,'' in \emph{USENIX Security Symposium (USENIX Security)}, 2022.

\bibitem{hybrid}
C.~Chen, R.~Kande, N.~Nguyen, F.~Andersen, A.~Tyagi, A.-R. Sadeghi, and J.~Rajendran, ``{HyPFuzz}: {Formal-Assisted} processor fuzzing,'' 2023, pp. 1361--1378.

\bibitem{Hur'21}
J.~Hur, S.~Song, D.~Kwon, E.~Baek, J.~Kim, and B.~Lee, ``{DifuzzRTL}: Differential fuzz testing to find {CPU} bugs,'' in \emph{IEEE Symposium on Security and Privacy (SP)}, 2021.

\bibitem{Canakci'23}
S.~Canakci, C.~Rajapaksha, L.~Delshadtehrani, A.~Nataraja, M.~Taylor, M.~Egele, and A.~Joshi, ``Processorfuzz: Processor fuzzing with control and status registers guidance,'' in \emph{IEEE International Symposium on Hardware Oriented Security and Trust (HOST)}, 2023.

\bibitem{RISC-V}
\BIBentryALTinterwordspacing
R.-V.~W. Group, ``Risc-v,'' last accessed: 11/18/2024. [Online]. Available: \url{https://riscv.org/}
\BIBentrySTDinterwordspacing

\bibitem{Morlkx}
\BIBentryALTinterwordspacing
OpenRISC, ``mor1kx - an openrisc processor {IP} core,'' last accessed: 11/18/2024. [Online]. Available: \url{https://github.com/openrisc/mor1kx}
\BIBentrySTDinterwordspacing

\bibitem{Laeufer'18}
K.~Laeufer, J.~Koenig, D.~Kim, J.~Bachrach, and K.~Sen, ``{RFUZZ}: Coverage-directed fuzz testing of rtl on fpgas,'' in \emph{IEEE/ACM International Conference on Computer-Aided Design (ICCAD)}, 2018.

\bibitem{direct}
S.~Canakci, L.~Delshadtehrani, F.~Eris, M.~B. Taylor, M.~Egele, and A.~Joshi, ``Directfuzz: Automated test generation for rtl designs using directed graybox fuzzing,'' in \emph{2021 58th ACM/IEEE Design Automation Conference (DAC)}, 2021, pp. 529--534.

\bibitem{saravanan2025intelligentgrayboxfuzzingatpgguided}
\BIBentryALTinterwordspacing
R.~Saravanan, S.~Paria, A.~Dasgupta, S.~Bhunia, and S.~M.~P. D, ``Intelligent graybox fuzzing via atpg-guided seed generation and submodule analysis,'' 2025. [Online]. Available: \url{https://arxiv.org/abs/2509.20808}
\BIBentrySTDinterwordspacing

\bibitem{Verilator}
\BIBentryALTinterwordspacing
Verilator, ``Welcome to verilator,'' last accessed: 11/18/2024. [Online]. Available: \url{https://www.veripool.org/verilator/}
\BIBentrySTDinterwordspacing

\bibitem{pico}
\BIBentryALTinterwordspacing
``Pico{RV}32,'' last accessed: 11/18/2024. [Online]. Available: \url{https://github.com/YosysHQ/picorv32}
\BIBentrySTDinterwordspacing

\bibitem{RISC}
\BIBentryALTinterwordspacing
O.~W. Group, ``Openrisc,'' last accessed: 11/18/2024. [Online]. Available: \url{https://openrisc.io/}
\BIBentrySTDinterwordspacing

\bibitem{Trippel'22}
T.~Trippel, K.~G. Shin, A.~Chernyakhovsky, G.~Kelly, D.~Rizzo, and M.~Hicks, ``Fuzzing hardware like software,'' in \emph{USENIX Security Symposium (USENIX Security)}, 2022.

\bibitem{borkar2024whisperfuzz}
P.~Borkar, C.~Chen, M.~Rostami, N.~Singh, R.~Kande, A.-R. Sadeghi, C.~Rebeiro, and J.~Rajendran, ``{WhisperFuzz: White-Box Fuzzing for Detecting and Locating Timing Vulnerabilities in Processors},'' \emph{arXiv preprint arXiv:2402.03704}, 2024.

\bibitem{hyperprop}
M.~R. Clarkson and F.~B. Schneider, ``Hyperproperties,'' in \emph{2008 21st IEEE Computer Security Foundations Symposium}, 2008, pp. 51--65.

\bibitem{det}
V.~Gohil, S.~Patnaik, H.~Guo, D.~Kalathil, and J.~Rajendran, ``Deterrent: Detecting trojans using reinforcement learning,'' \emph{IEEE Transactions on Computer-Aided Design of Integrated Circuits and Systems}, vol.~43, no.~1, pp. 57--70, 2024.

\bibitem{Huang2016MERSST}
Y.~Huang, S.~Bhunia, and P.~Mishra, ``Mers: Statistical test generation for side-channel analysis based trojan detection,'' \emph{Proceedings of the 2016 ACM SIGSAC Conference on Computer and Communications Security}, 2016.

\bibitem{scop}
Z.~Pan and P.~Mishra, ``Automated test generation for hardware trojan detection using reinforcement learning,'' in \emph{2021 26th Asia and South Pacific Design Automation Conference (ASP-DAC)}, 2021, pp. 408--413.

\bibitem{Ahmed'18}
A.~Ahmed, F.~Farahmandi, and P.~Mishra, ``Directed test generation using concolic testing on {RTL} models,'' in \emph{Design, Automation \& Test in Europe Conference \& Exhibition (DATE)}, 2018.

\bibitem{Farahmandi}
F.~Farahmandi, Y.~Huang, and P.~Mishra, ``Trojan localization using symbolic algebra,'' in \emph{2017 22nd Asia and South Pacific Design Automation Conference (ASP-DAC)}, 2017, pp. 591--597.

\bibitem{Meng'22}
X.~Meng, S.~Kundu, A.~K. Kanuparthi, and K.~Basu, ``{RTL}-contest: Concolic testing on {RTL} for detecting security vulnerabilities,'' \emph{IEEE Transactions on Computer-Aided Design of Integrated Circuits and Systems}, vol.~41, no.~3, pp. 466--477, 2022.

\bibitem{Shakya2017BenchmarkingOH}
B.~Shakya, M.~T. He, H.~Salmani, D.~Forte, S.~Bhunia, and M.~M. Tehranipoor, ``Benchmarking of hardware trojans and maliciously affected circuits,'' \emph{Journal of Hardware and Systems Security}, vol.~1, pp. 85 -- 102, 2017.

\bibitem{opentitan}
\BIBentryALTinterwordspacing
Google, ``Opentitan,'' last Accessed : 6/1/2024. [Online]. Available: \url{https://opentitan.org/}
\BIBentrySTDinterwordspacing

\bibitem{ibex}
\BIBentryALTinterwordspacing
lowRISC, ``Ibex,'' last Accessed : 6/1/2024. [Online]. Available: \url{https://lowrisc.org/}
\BIBentrySTDinterwordspacing

\bibitem{open}
\BIBentryALTinterwordspacing
OpenCores, ``Opencores,'' last Accessed : 6/1/2024. [Online]. Available: \url{https://opencores.org/}
\BIBentrySTDinterwordspacing

\end{thebibliography}

\vspace{12pt}

\end{document}